\documentclass[aps,pre,showpacs,preprint,superscriptaddress]{revtex4}
\usepackage{graphicx}
\usepackage{amsmath}
\usepackage{ulem}
\usepackage{bm}

\begin{document}

\title{Effects of ion mobility and positron fraction on solitary waves in weak relativistic electron-positron-ion plasma}
\author{Ding Lu}
\author{Zi-Liang Li}
\author{Bai-Song Xie\footnote{Corresponding author. Email address: bsxie@bnu.edu.cn}}
\affiliation{Key Laboratory of Beam Technology and Materials Modification of the Ministry of Education, and College
of Nuclear Science and Technology, Beijing Normal University, Beijing 100875, China}

\begin{abstract}
Effects of ion mobility and positron fraction on solitary waves of envelop of laser field and
potential of electrostatic field  in weak relativistic electron-positron-ion plasma are investigated.
The parameter region for the existence of solitary waves is obtained analytically, and the reasonable
choice of parameters is clarified. Both cases of mobile and immobile ions are considered. It is found that
the amplitudes of solitary waves in the former case are larger compared to the latter case. For small plasma density, the localized solitary wave solutions in terms of approximate perturbation analytical method are consistent well with that
by exact numerical calculations. However as the plasma density increases the analytical method loses its validity
more and more. The influence of the positron fraction on the amplitudes of solitary waves shows a monotonous
increasing relation. Implication of our results to the particle acceleration is also discussed briefly.
\end{abstract}

\maketitle

\section{Introduction}

Since the relativistic laser-plasma interaction was first investigated by Akhiezer and Polovin \cite{Pol}, many nonlinear
phenomena have been found, such as solitons, vortices, double layers and so on. In particular the discovery of electromagnetic solitary waves have attracted the attention of many people due to its robust and resilient behaviors \cite{Whit}. In past decades many works have been performed in conventional electron-ion (EI) plasma \cite{Bulanov,Berezhiani1,D.Farina,Xie,Xie1}.
On the other hand, there exists also electron-positron-ion (EPI) plasma in most astrophysical environments and in the laboratory \cite{Chen}, for example, in the pulsar magnetosphere \cite{Michel}, the active galactic nuclei \cite{Begelman}, and the early universe \cite{Sturrock}. Recently the studies of nonlinear waves in the intense laser field interacting with EPI plasma have revealed some new features \cite{Berezhiani,Bere,Berezhiani2,Popel,Shukla,Haque,Kakati,Mishra}. Usually the EPI plasma is very different from the conventional EI plasma and also different from the pure electron-positron (EP) plasma \cite{Bere}. An obvious fact is that there is no soliton solutions for pure EP plasma because the charge density cancels each other for charge neutralized electrons and positrons.

Let us recall some researches that have been done about the localized solitary wave solutions in two-component plasma, EI or/and EP, and three-component plasma, EPI. Farina and Bulanov \cite{D.Farina,Farina} studied the relativistic electromagnetic solitons in EI and EP plasma. In our previous study we obtained the parameter region of existence of solitons and the bifurcation diagram for EI plasma \cite{Hua}. Berezhiani {\it et al.} \cite{Berezhiani1} studied the relativistic solitary waves in cold EP plasma in an external magnetic field. They also found the large amplitude envelope solitons analytically in EPI plasma \cite{Berezhiani2}.
By using the reductive perturbation method, Mahmood {\it et al.} \cite{Mahmood} obtained the small amplitude Korteweg-de Vries (KdV) soliton solutions in hot EPI plasma and they also considered the influence of temperature ratio of  electrons to positrons on solitons. Lehmann {\it et al.} \cite{Lehmann} studied the manifolds of periodic solutions by Poincar\'{e} section plots, they found that the ion dynamic strongly affect the structure of the phase plot and the ion motion must not be neglected in the high-intensity regime. Some other works are also done, such as the studies of ion-arouse solitary waves \cite{Popel} and vortices \cite{Shukla,Haque} in EPI plasma.

The propose of this paper is to investigate solitary waves in the cold and unmagnetized EPI plasma. We will focus on three important aspects. First we have to choose a set of parameters which are relatively practical and physical by the requirement of approximated wave dispersion. It is noted that this point is always either omitted or less discussed in many publications before.
Second we employ the numerical technique to solve the coupling equations of envelop of laser field and potential of plasma electrostatic field, which can provide an exact solitary wave solution. Third we generalize the approximate perturbation analytic method developed by others \cite{Bere} for immobile ion case to the case including the ion motion. Motivated by these factors, we study both cases of mobile and immobile ions. The numerical results are compared to the analytical ones. We also hope that our results in this paper are helpful to clarify the properties of nonlinear solitary waves in EPI plasma. Before giving the exact numerical solutions, we have made a detailed stability analysis for fixed point of the system and the parameter region for the existence of solitary waves, like in Ref. \cite{Hua}, which can give relatively a deeper understanding of nonlinear coherent structure in multi-component plasma.

The paper is organized as follows. Section II reviews the theoretical model of laser-plasma interaction for completeness of paper, and presents the parameter region that solitary waves may exist analytically in terms of nonlinear dynamics method. Section III exhibits the exact numerical solutions in both cases of mobile and immobile ions. And the numerical results are compared to the approximate analytical solutions. A conclusion is given in Sec. IV. Finally the derivation of analytical method by perturbation series expansion technique generalized to the mobile ion case is given in Appendix.

\section{Theoretical model and basic formalism}

The basic formalism of studied problem is certainly from the Maxwell equations and hydrodynamic equations. In the following the involving physical quantities are normalized as time $\omega_\mathrm{p}t$, length
$k_\mathrm{p}x$, potential $e\phi/m_ec^2$, electromagnetic fields
$eE/m_ec\omega_\mathrm{p}$ and $eB/m_ec\omega_\mathrm{p}$, density
$n/n_{0e}$, velocity $v/c$ and momentum $p/m_\alpha c$, respectively, where
$\omega_\mathrm{p}$ is the electron plasma frequency, $m_e$ is the electron rest mass and $n_{0e}$ is the unperturbed electron density. Charges $q_\mathrm{\alpha}$ are normalized to $-e$, where $\alpha$ indicates the particle species ($\alpha=e,p,i$), so for electron $q_\mathrm{e}=1$, for positron and ion $q_\mathrm{p}= q_\mathrm{i}=-1$,  $\rho_\alpha$ is defined as $m_e/m_\alpha$, so for electron and positron, $\rho_e=\rho_p=1$, and for ion (for simplicity we assumed it as proton) $\rho_\mathrm{i}= 1/1836$. We shall consider that the system is initially an equilibrium state which is characterized by $n_\mathrm{0e}=n_\mathrm{0p}+n_\mathrm{0i}$ to ensure the charge neutrality. The positron fraction can be defined as $\chi=n_\mathrm{0p}/n_\mathrm{0e}$, which is the ratio of positron to electron initially.

By using the normalized quantities mentioned above,  the Maxwell equations and hydrodynamic equations are written in the form as
\begin{eqnarray}
&&\nabla^2\bm{A}-\frac{\partial^2}{\partial t^2}\bm{A}-\frac{\partial}{\partial t}\nabla \phi= n_e \bm{v}_e-n_p \bm{v}_p-n_i \bm{v}_i , \label{eq1}\\
&&\nabla^2 \phi=n_e-n_p-n_i  , \label{eq2}\\
&&\frac{\partial n_\alpha}{\partial t}+ \nabla\cdot(n_\alpha \bm{v}_\alpha)=0  , \label{eq3}\\
&&\frac{\partial \bm{P}_\alpha}{\partial t}=\nabla(\rho_\alpha q_\alpha \phi+\gamma_\alpha)+\bm{v}_\alpha\times\nabla\times \bm{P}_\alpha \label{eq4}
\end{eqnarray}
where $\bm{P}_\alpha=\bm{p}_\alpha - \rho_\alpha q_\alpha\bm{A}$ is the canonical momentum, $\gamma_\alpha=\sqrt{1+|\bm{p}_\alpha|^2}$ is the relativistic factor and $\bm{v}_\alpha=\bm{p}_\alpha/\gamma_\alpha$ is particle fluid velocity. Note that we have chosen the Coulomb gauge $\nabla\cdot \bm{A}=0 $.

As in most studies before, for convenience, we just consider the problem that all the quantities vary only along the direction of propagation $x$. We look for the solution of vector potential as the form of $\bm{A}=a(\xi)\exp(i\omega \tau)$, where the new variables $\xi = x-vt$, $\tau=t-vx$ are introduced and $v$ is the laser group velocity. By the way the wave traveling phase velocity is just $1/v$.

In the quasistatic approximation and the initial plasma conditions of $p_\alpha=0, n_e=1, n_p=\chi, n_i=1-\chi$ at infinity, then we get the reduced coupling  equations as follows
\begin{eqnarray}
\frac{d^2a}{d\xi^2}&=&-\omega^2a+{a\mathcal{F}_1}/{\varepsilon^2}\nonumber \\
&\equiv&g(a,\phi),\label{eq5}\\
\frac{d^2\phi}{d\xi^2}&=&{\mathcal{F}_2}/{\varepsilon^2}\nonumber \\
&\equiv&h(a,\phi).\label{eq6}
\end{eqnarray}
where $\varepsilon^2=1-v^2$ (obviously $0 \leq \varepsilon \leq1$) and two auxiliary quantities are introduced as
\begin{eqnarray}
\mathcal{F}_1=\sqrt{1-\varepsilon^2}\bigg[\frac{1}{R_e}+\frac{\chi}{R_p}+\frac{\rho_i(1-\chi)}{R_i}\bigg],\label{eq7}\\
\mathcal{F}_2=\sqrt{1-\varepsilon^2}\bigg[\frac{\psi_e}{R_e}-\frac{\chi\psi_p}{R_p}
-\frac{(1-\chi)\psi_i}{R_i} \bigg]. \label{eq8}
\end{eqnarray}
with $\psi_\alpha=\rho_\alpha q_\alpha\phi+1$ and $R_\alpha=\sqrt{\psi_\alpha^2-(1-v^2)(1+\rho_\alpha^2a^2)}$.
Now Eqs.(\ref{eq5}) and (\ref{eq6}) constitute a set of coupled ordinary differential equations (ODEs) for the vector potential of the laser pulse and wake potential of plasma, which are our starting point for further investigations in the following.

We shall first discuss the nonlinear dynamics that is hidden in these equations. In order to predict the existence of the solitary waves, Eqs.(\ref{eq5}) and (\ref{eq6}) can be reduced to a set of one-order ODEs as $a'=b$, $b'=g(a,\phi)$, $\phi'=c$, $c'=h(a,\phi)$, where the prime denotes as $d/d\xi$. The fixed points can be calculated from $b=0$, $g(a,\phi)=0$, $c=0$ and $h(a,\phi)=0$. For example, in the case of mobile ion the fixed point $(a, a', \phi, \phi' )=(0,0,0,0)$ is found.
It is easy to find that the characteristic roots of Jacobi determinant are $ \lambda_{1,2}^2= [1+\chi+\rho_i (1-\chi)]/\varepsilon^2-\omega^2$ and $\lambda_{3,4}^2= -[1+\chi+\rho_i (1-\chi)]/(1-\varepsilon^2) $. Let us denote  $\delta=\varepsilon \omega$. Obviously if $ 0 < \delta < \delta_c=\sqrt{1+\chi+\rho_i (1-\chi)}$, the characteristic roots are
$\lambda_{1,2}=\pm  \sqrt{[1+\chi+\rho_i (1-\chi)]/\varepsilon^2-\omega^2}$ and
$\lambda_{3,4}= \pm i \sqrt{[1+\chi+\rho_i (1-\chi)]/(1-\varepsilon^2)} $. Accordingly the stability analysis mentioned above shows that the fixed point $(0,0,0,0)$ is a saddle-center when $0 <\delta<\delta_c$. This means that in this region there exist solitary waves.

Because ions are much heavier than the electrons, so the ion motion could be neglected for simplified treatment in some situations. Thus for immobile ion case, similar to the mobile ion case, the coupling equations are in the same form except $ \rho_i = 0$, and with
\begin{eqnarray}
&&\mathcal{F}_3=\sqrt{1-\varepsilon^2}\bigg[\frac{1}{R_e}+\frac{\chi}{R_p}\bigg],\label{eq9}\\
&&\mathcal{F}_4=\sqrt{1-\varepsilon^2}\bigg[\frac{\psi_e}{R_e}-\frac{\chi\psi_p}{R_p}
-\frac{(1-\chi)}{\sqrt{1-\varepsilon^2}} \bigg], \label{eq10}
\end{eqnarray}
replacing the former $\mathcal{F}_1$ and $\mathcal{F}_2$. Similarly we find that the fixed point is also $(0, 0, 0, 0)$ and the stability analysis shows that it is also the saddle-center. The solitary waves can be available in the region of $0 < \delta<\delta_c= \sqrt{1+\chi}$.

\begin{figure}[htbp]\suppressfloats
\includegraphics[width=15cm]{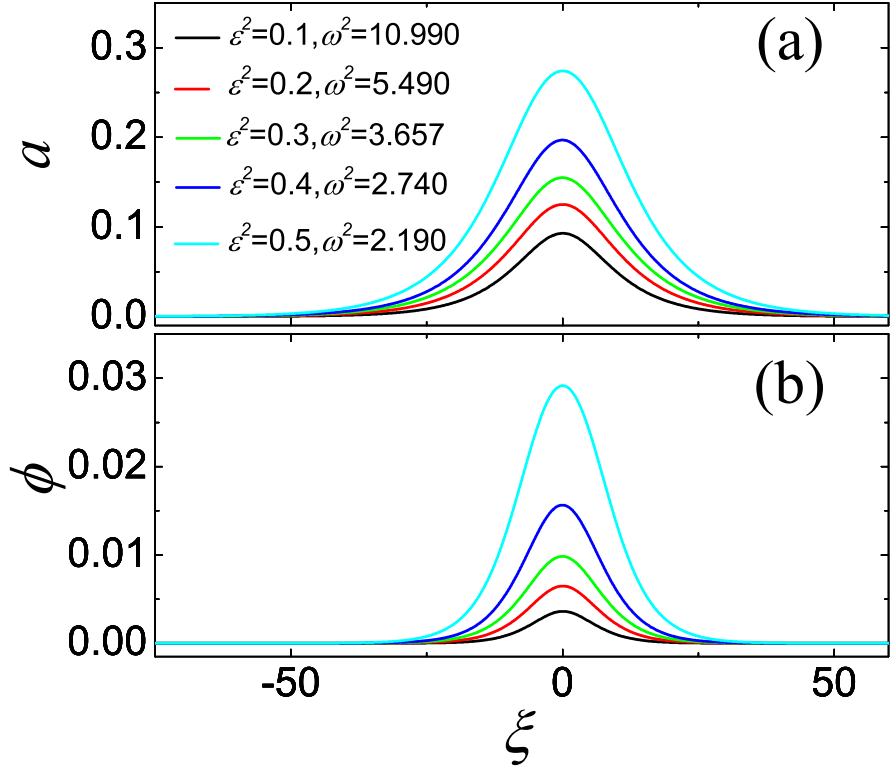}
\caption{\label{Fig1}(Color online) Solitary wave solutions of vector potential $a$ in (a) and electrostatic potential $\phi$ in (b) for mobile ion case. The positron fraction is $\chi=0.1$ and the other parameters are seen in text.}
\end{figure}

\section{Results of localized solitary wave solutions}

Before giving numerical and analytical solutions in the following, it is necessary to see how to choose the system parameters which are important for practical physical system. It is well known that in non-relativistic frame we have the normalized laser group velocity $v=\sqrt{(1-\omega^{-2})}$ for conventional EI plasma (note that the normalized laser frequency is the ratio of real laser frequency $\omega_L$ to equilibrium plasma frequency $\omega_p$, i.e. $\omega=\omega_L/\omega_p$). This relation comes from the dispersion of transverse electromagnetic waves propagating in plasma. A further relation is that $\varepsilon^2=1-v^2=1/\omega^2$ which will lead to $\delta=\varepsilon \omega=1$. However in the weak relativistic frame due to the plasma mass is modified by the relativistic factor so that $\delta \approx 1/\gamma_e \lesssim1$. Similarly for EIP plasma we have $\delta \approx \delta_c/\gamma_e \lesssim\delta_c$ for both of mobile and immobile ion cases. These discussions lead to two facts. One is that indeed the practical laser and plasma parameters can lie in the region where the solitary waves can exist analyzed in the last section. The other, maybe more important, is that the production quantity $\delta=\varepsilon \omega$ should be near to the value of bifurcation point $\delta_c$ because the studied problem is the weak relativistic in which the relativistical factor $\gamma_e$ is a little larger than $1$. In fact our studies below support and confirm these analysis. For convenience we should remind that the larger parameter $\varepsilon^2$ ($\approx n/n_c$) corresponds to  the higher plasma density but the smaller laser group velocity.

\subsection{Effects of ion mobility}

The 4th order Rung-Kutta method is employed in our numerical techniques. We can find a series of numerical solitary wave
solutions for Eqs.(\ref{eq5}) and (\ref{eq6}) with different parameters of $ \varepsilon $ and $\omega $ if $\chi$ is fixed. For illustration some typical numerical results are shown in Fig.\ref{Fig1} in mobile ion case. Specifically, Fig.\ref{Fig1}(a) is the envelop of vector potential $a$ and Fig.\ref{Fig1}(b) is the electrostatic potential $\phi$. We have chosen $\chi=0.1$ and five sets of $\varepsilon$ and $\omega$ as $(\varepsilon^2,\omega^2)$ =$(0.1, 10.990)$ (black line), $(0.2, 5.490)$ (red line), $(0.3, 3.657)$ (green line), $(0.4, 2.740)$ (blue line) and $(0.5, 2.190)$ (cyan line). From the discussion mentioned above these parameters are chosen to near the bifurcation point, i.e., $\delta<\delta_c$  but $\delta_c-\delta \ll 1$. One can clearly see that with the increase of plasma density the amplitudes of solitary waves are increased accordingly in the mobile ion case. For immobile ion case, we can get the corresponding solitary wave solutions in the same five sets of $\varepsilon$ and $\omega$, and the amplitudes of solitary waves increase also with the plasma density.

\begin{figure}[htbp]\suppressfloats
\includegraphics[width=15cm]{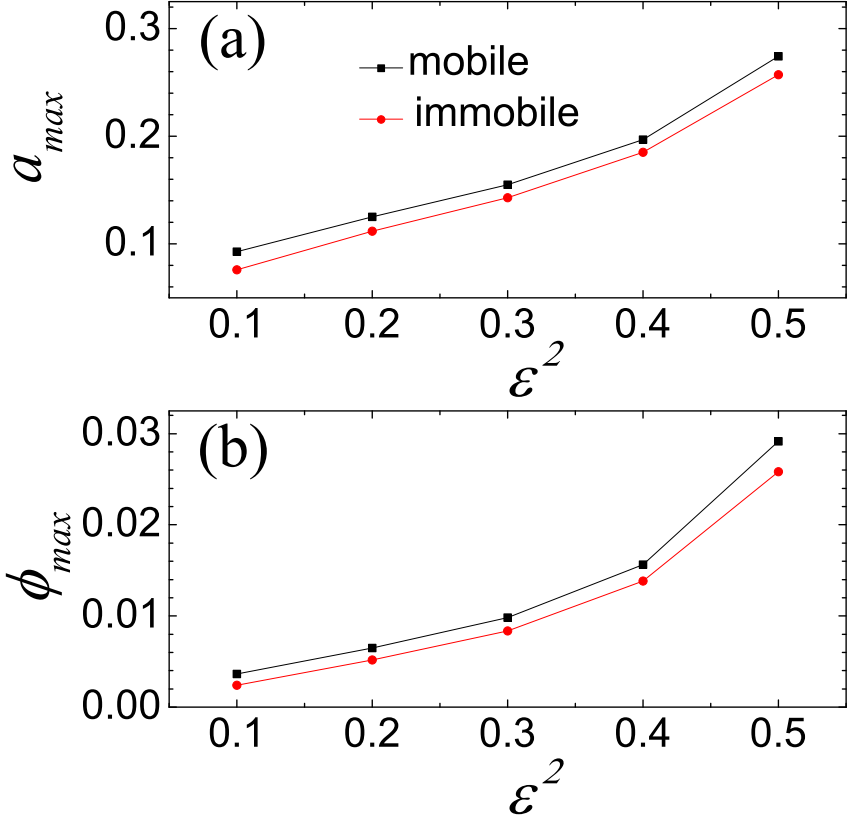}
\caption{\label{Fig2} (Color online) Maximum values of vector potential $a_{max}$ and scalar potential $\phi_{max}$
in mobile (black squares)  and immobile (red circles) ion cases.}
\end{figure}

For comparisons, the maximum values of vector potential and scalar potential in both of mobile (black squares) and
immobile (red circles) ion cases are plotted in Fig.\ref{Fig2}. It is found that the solitary wave amplitudes of vector potential $a$ and electrostatic potential $\phi$ in the mobile ion case are always larger than that in the immobile ion case under the
same conditions of $\varepsilon$, $\omega$, and $\chi$. It is not surprising because only stronger laser can excite the larger
plasma motion, which intrigues the larger plasma potential. This self-consistent physical picture is a typical nonlinear
characteristic in laser-plasma interaction which is associated strongly to the ponderomotive of laser field as well as the
electrostatic force due to plasma charge density.

\subsection{Comparison of numerical and analytical results}

In order to compare the numerical solutions to analytical results the derivation of analytical solutions is given in Appendix.
We get the concrete analytical expression for solitary wave solutions as Eq.(\ref{eqa17}) by using a similar perturbation series
expansion method which has been developed by many people \cite{varma,Rao,N.N,Mofiz,Bere} but the scheme and results
are generalized to include the ion dynamics.

\begin{figure}[htbp]\suppressfloats
\includegraphics[width=15cm]{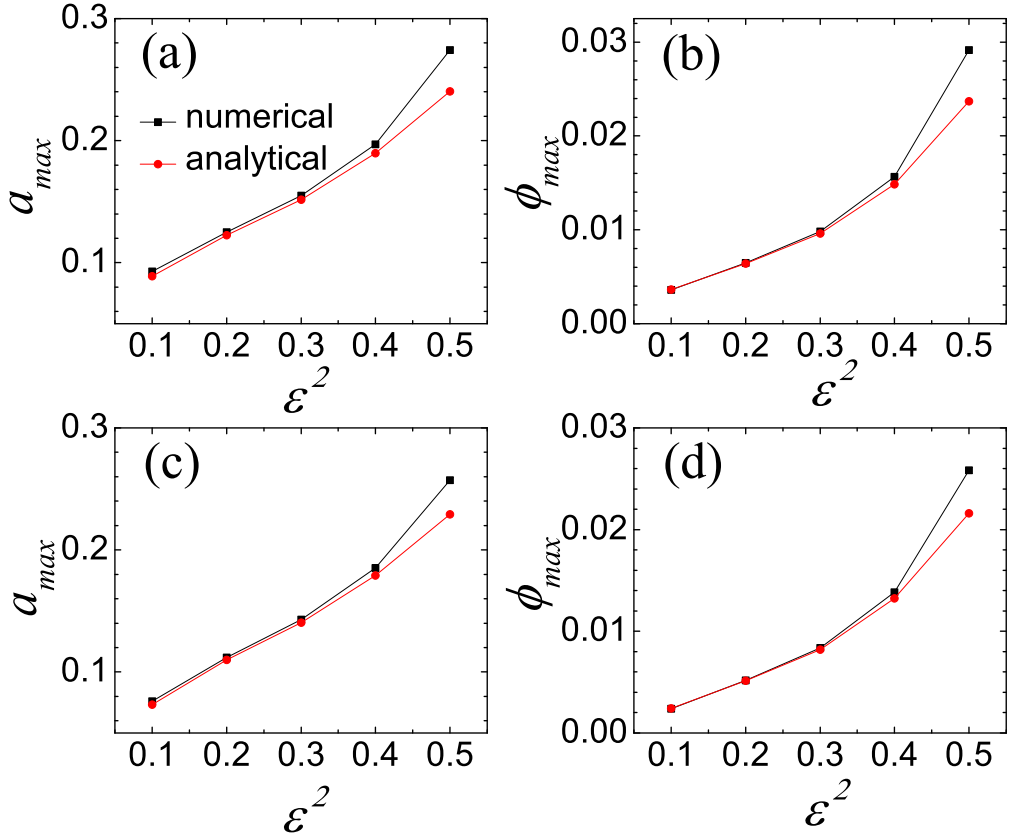}
\caption{\label{Fig3} (Color online) Maximum values of vector potential $a_{max}$ and scalar potential $\phi_{max}$ for  mobile ion case (upper row) and immobile ion case (lower row) by the numerical results (black squares) and analytical results (red circles).}
\end{figure}

Figure \ref{Fig3} demonstrates the maximum values of vector potential $a_{max}$ and scalar potential $\phi_{max}$ in both of mobile (upper row) and immobile (lower row) ion cases with the numerical results (black square) and analytical results (red circles). The choice of system parameters is the same as in Fig.\ref{Fig1}. One can find that as the increase of the plasma density, $\varepsilon^2$, the amplitudes of solitary waves $a$ and $\phi$ are increasing monotonously too. Meanwhile the analytical results agree with the numerical results very well for small values of $\varepsilon^2$, i.e. relatively tenuous plasma. However, as the plasma becomes dense enough, the underestimation of analytical solutions becomes more and more severe.

\begin{figure}[htbp]\suppressfloats
\includegraphics[width=15cm]{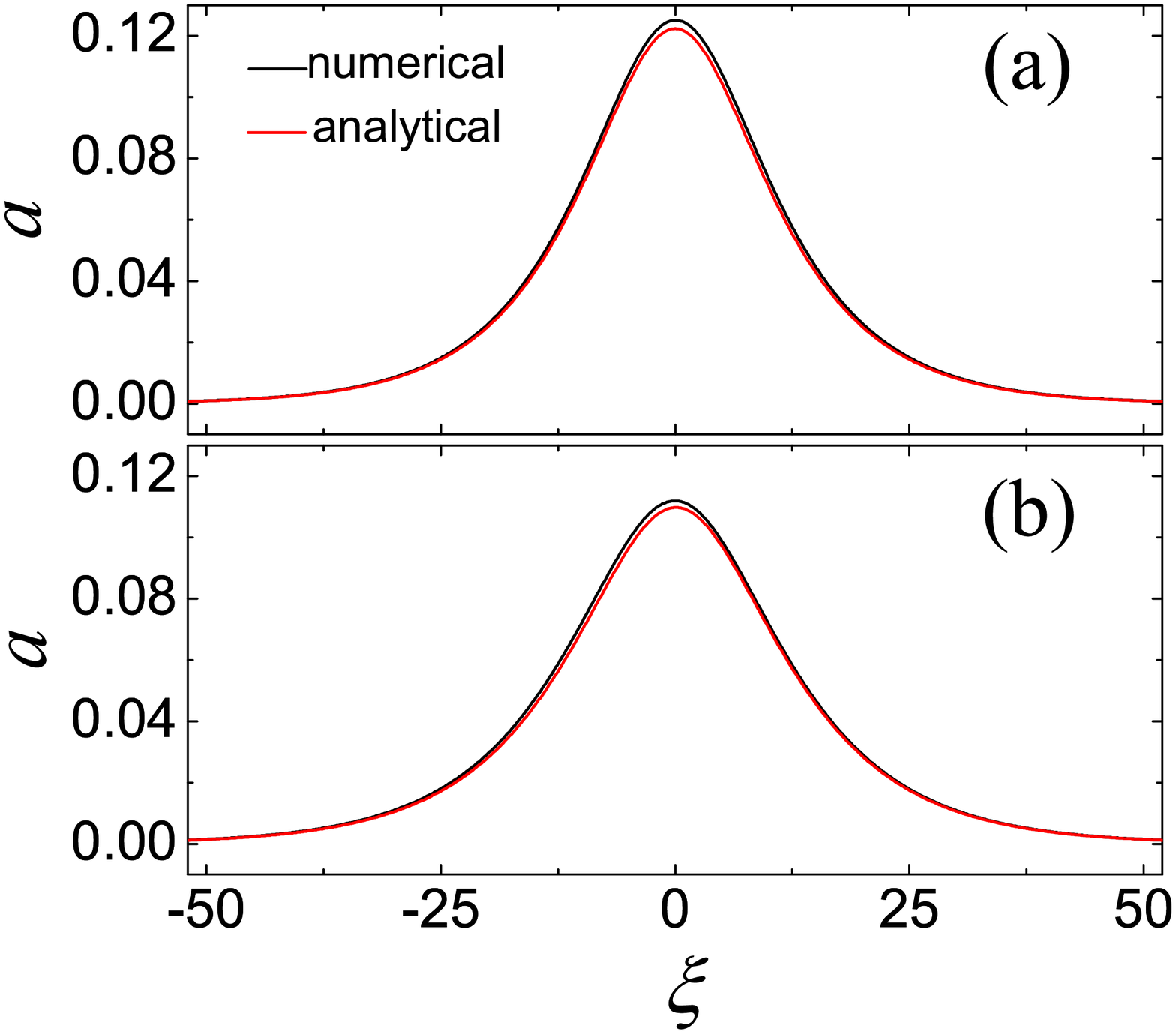}
\caption{\label{Fig4}(Color online) The numerical and analytical solutions of vector potential $a$  for $\chi=0.1$,  $\varepsilon^2=0.2$, and $\omega^2=5.49$ in both of mobile ion case (upper row) and immobile ion case (lower row). }
\end{figure}

\begin{figure}[htbp]\suppressfloats
\includegraphics[width=15cm]{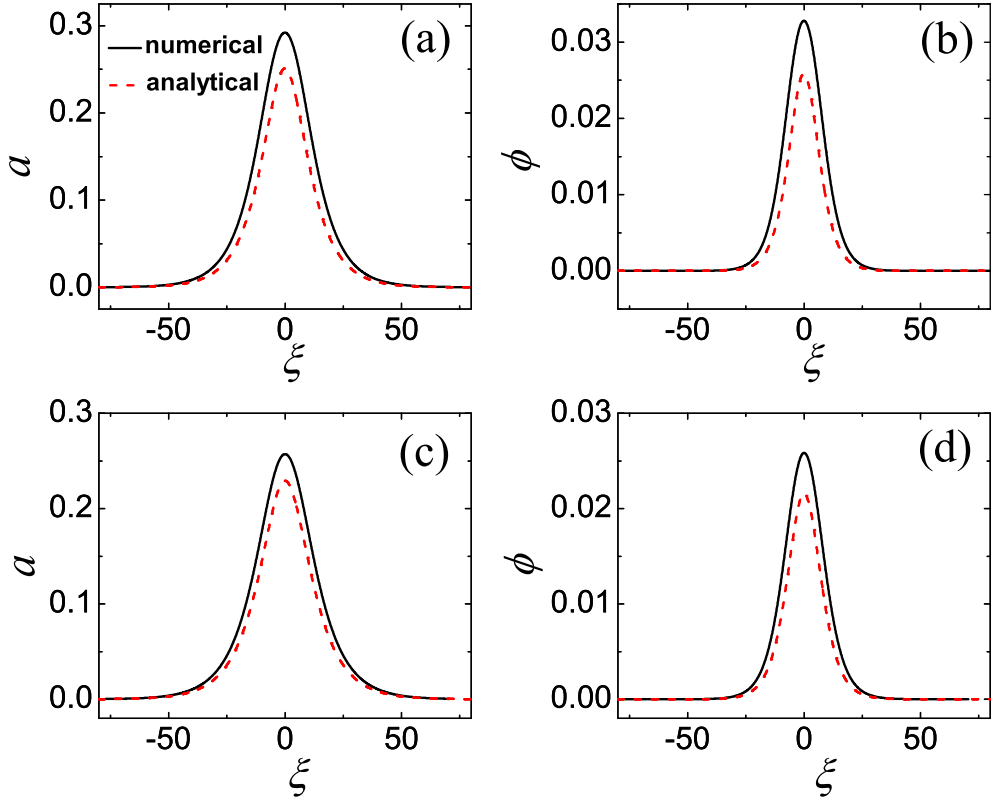}
\caption{\label{Fig5} (Color online) The numerical and analytical solutions of vector potential $a$ and electrostatic potential $\phi$ for $\varepsilon^2=0.5$ and $\omega^2=2.19$ in both of mobile ion case (upper row) and immobile ion case (lower row).}
\end{figure}

For the case of smaller plasma density, $\varepsilon^2= 0.2$, and when parameters $\chi= 0.1$ and $\omega^2= 5.49$ are given,
the numerical and analytical solutions of vector potential $a$ in both mobile ion case (upper row) and immobile ion case (lower row) are shown
in Fig.\ref{Fig4}. The numerical solutions and the analytical ones are plotted by black solid line and red solid line, respectively. Indeed the numerical results and analytical ones are almost the same because the analytical expression is valid well for the choice of parameters.
On the other hand for the higher plasma density, $\varepsilon^2= 0.5$, and given the other parameters as $\chi= 0.1$ and
$\omega^2= 2.19$, the corresponding numerical solutions and analytical solutions in both of mobile ion case (upper row)
and immobile ion case (lower row) are shown in Fig.\ref{Fig5}. Obviously there are remarkable discrepancy between
the numerical and analytical results. We think it attributes to a fact that the series expansion terms are not enough.
If we take higher orders of terms in Eq.(\ref{eqa4}), for example $n = 3$ or $n = 4$, the analytical solutions will become
more accurate which will be in accordance with the numerical ones. These observations indicate that the
limiting condition Eq.(\ref{eqa19}) in analytical method plays an important role. The series expansion
would lose its validity rapidly if the convergence condition is violated, e.g., for the higher plasma density.
Therefore the numerical solving scheme, presented in this paper, is necessary due to its
exactness and simplicity compared to the analytical method.

\subsection{Effects of positron fraction}

\begin{figure}[htbp]\suppressfloats
\includegraphics[width=15cm]{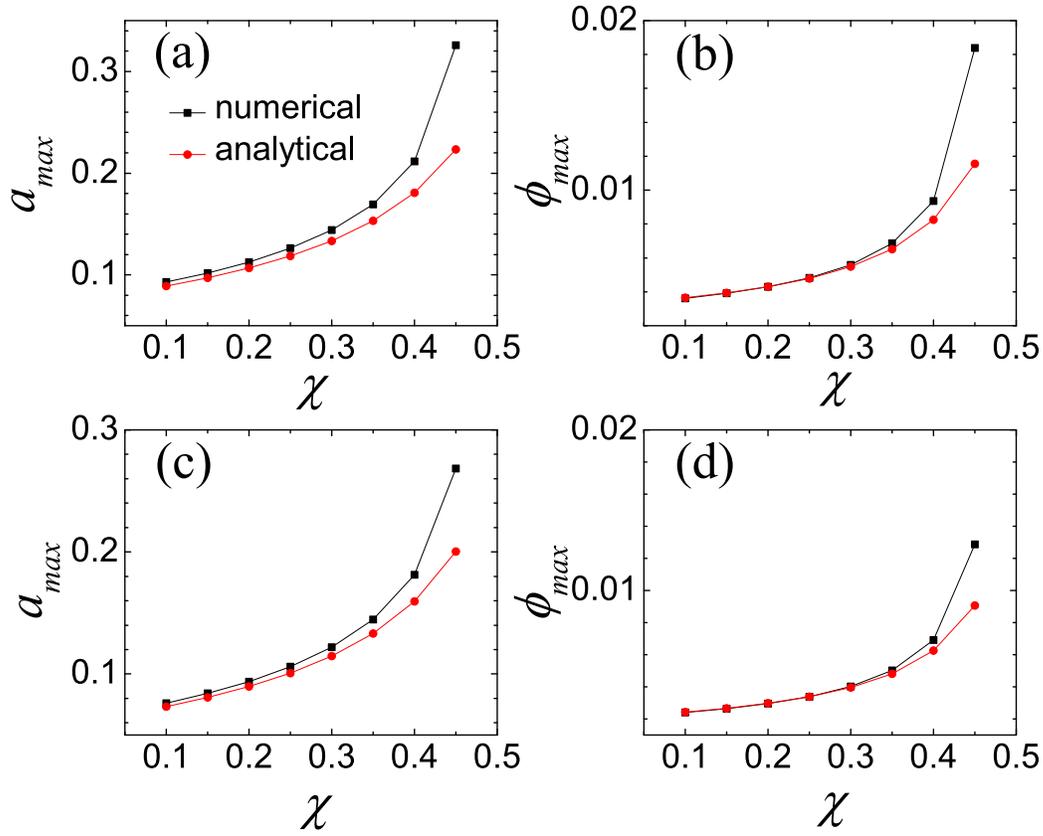}
\caption{\label{Fig6} (Color online) The numerical (black squares) and analytical (red circles) results of vector potential maximum values $a_{max}$  and scalar potential maximum values $\phi_{max}$ changing with $\chi$ for both of the cases of mobile ion (upper row) and immobile ion case (lower row) with parameter $\varepsilon^2=0.1$.}
\end{figure}

To see the effect of positron fraction on solitary waves we plot in Fig.\ref{Fig6} the maximum values of vector potential
$a_{max}$  and scalar potential $\phi_{max}$ for mobile ion case (upper row) and immobile ion case (lower row) with
the numerical results (black square) and analytical results (red circles). One can find that as increase of $\chi$ both values
of $a_{max}$ and $\phi_{max}$ are increased monotonously, i.e., the ratio of positron to electron density can enhance
the amplitudes of solitary waves. We also find the analytical results can fit the numerical results very well with small values
of $\chi$ but deviate more and more as $\chi$ becomes large. Especially when $\chi$ lies in the region of $(1- \chi )\ll 1$ (see Appendix), the analytical method will be invalid severely. Furthermore, by comparing the values of $a_{max}$ and $\phi_{max}$ in Fig.\ref{Fig6}, we can find that the maximum values of $a$ and $\phi$ in mobile ion case is larger
than that in immobile ion case for the same parameters. It is just the effects of ion dynamics on the solitary wave solutions.

\section{Conclusions}

In a summary, in this paper, the solitary waves have been found numerically and analytically in electron-positron-ion plasma
with the condition of charge neutrality initially and we have considered both cases of mobile and immobile ions.
The existence of solitary wave solutions have been found analytically in a certain parameter region. We also find that
when ion is mobile, compared to immobile ion, both the parameter region and the amplitudes of solitary waves become larger. In addition, the positron concentration has a significant impact on the solitary waves in the considered plasma system. It is found that as increase of positron fraction, the parameter region of solitary waves becomes larger and the amplitudes of the waves too. We also compare the numerical results with  analytical solutions, and they
are almost the same except for the cases of high plasma density or/and high positron fraction.

It is worthy to note that our research may be helpful to understand charged particle acceleration in plasma, for example, the electrons can be accelerated from the bottom to reach the top in wake plasma potential. The coupling between laser field $a$ and wake field $\phi$ leads to nonlinear localized structure like solitary waves presented in this paper would not only make the effective charged particle acceleration possible but also open a possible new way to increase the acceleration efficiency through adjusting the system parameters, e.g., the plasma density and the positron fraction, however, which beyond the scope of present paper and need to study in future in detail.

\begin{acknowledgments}
This work was supported by the National Natural Science Foundation of China (NNSFC) under the grant No.11175023 and No. 11335013 and partially by the Fundamental Research Funds for the Central
Universities (FRFCU). The computation was carried out at the HSCC
of the Beijing Normal University.
\end{acknowledgments}

\begin{appendix}
\section{ The derivation of analytical solution}
\renewcommand{\theequation}{A\arabic{equation}}
By employing the series expansion method which was presented in Ref.\cite{Bere}, we calculate the analytical solutions of Eq.(\ref{eq5}) and Eq.(\ref{eq6}). In this conservative system, it is not hard to find the integral of motion as
\begin{eqnarray}
\bigg(\frac{da}{d\xi}\bigg)^2-\frac{1}{\mathcal{R}}\bigg(\frac{d\phi}{d\xi}\bigg)^2=&& -\Omega \mathcal{A}-\frac{2v}{\mathcal{R}^2}\bigg(\sqrt{(1+\phi)^2-\mathcal{R}(1+\mathcal{A})}
+\chi(\sqrt{(1-\phi)^2-\mathcal{R}(1+\mathcal{A})} \nonumber\\
&&+\frac{1-\chi}{\rho_i}\sqrt{(1-\rho_i\phi)^2-\mathcal{R}(1+\rho_i^2\mathcal{A})}\bigg) +E=\mathcal{H}(\phi,\mathcal{A}),  \label{eqa1}
\end{eqnarray}
where $\mathcal{A}=a^2,\Omega=\omega^2, \mathcal{R}=\varepsilon^2$ and
\begin{eqnarray}
E=\frac{2v^2}{\mathcal{R}^2}(1+\chi+\frac{1-\chi}{\rho_i})  \label{eqa2}
\end{eqnarray}
is the integration constant which can be calculated from the boundary condition.

By using the energy integral and eliminating the independent variable $\xi$ in Eq.(\ref{eq5}) and Eq.(\ref{eq6}), \textbf{we }can get the following equation:
\begin{eqnarray}
4\mathcal{A} \mathcal{H}(\phi,\mathcal{A})\frac{d^2\phi}{d\mathcal{A}^2}-\frac{8}{\mathcal{R}}g(\phi,\mathcal{A})
\mathcal{A}\left(\frac{d\phi}{d\mathcal{A}}\right)^3+\frac{4}{\mathcal{R}}\mathcal{A}h(\phi,\mathcal{A})
\left(\frac{d\phi}{d\mathcal{A}}\right)^2   \nonumber\\
+2[g(\phi,\mathcal{A})+\mathcal{H}(\phi,\mathcal{A})] \frac{d\phi}{d\mathcal{A}}-h(\phi,\mathcal{A})=0  \label{eqa3}
\end{eqnarray}

We assume that the electrostatic field $\phi $ is the function of field amplitude $\mathcal{A}$ and use the series expansions as
\begin{eqnarray}
&&\phi(\mathcal{A})= \sum_n c_n \mathcal{A}^n,      \nonumber\\
&&(1-\phi)^2-\mathcal{R} (1+\mathcal{A})=\sum_n a_n \mathcal{A}^n,     \nonumber\\
&&(1+\phi)^2-\mathcal{R} (1+\mathcal{A})=\sum_n b_n \mathcal{A}^n,      \nonumber\\
&&(1-\rho_i \phi)^2-\mathcal{R} (1+\rho_i^2 \mathcal{A})=\sum_n d_n \mathcal{A}^n.    \label{eqa4}
\end{eqnarray}
with $c_0=0,a_0=b_0=d_0=1-\mathcal{R}$. For $n\geq1$, $a_n$, $b_n$ and $d_n$ can be expressed by $c_n$ as
\begin{eqnarray}
&&a_1=-2c_1-\mathcal{R}, a_2=c_1^2-2c_2, a_3=2c_1c_2-2c_3, ... \nonumber\\
&&b_1=2c_1-\mathcal{R}, b_2=c_1^2+2c_2, b_3=2c_1c_2+2c_3, ... \nonumber\\
&&d_1=-2\rho_ic_1-\rho_i^2\mathcal{R}, d_2=\rho_i^2c_1^2-2\rho_ic_2, d_3=2\rho_i^2c_1c_2-2\rho_ic_3, ... \label{eqa5}
\end{eqnarray}
We also assume the following expansions are valid
\begin{equation}
[(1-\phi)^2-\mathcal{R}(1+\mathcal{A})]^{\pm\frac{1}{2}}= v^{\pm 1}(1\pm\frac{1}{2 v^2}\sum_n a_n \mathcal{A}^n),\nonumber
\end{equation}
\begin{equation}
[(1+\phi)^2-\mathcal{R} (1+\mathcal{A})]^{\pm\frac{1}{2}}= v^{\pm 1}(1\pm\frac{1}{2 v^2}\sum_n b_n \mathcal{A}^n) ,\label{eqa6}
\end{equation}
\begin{equation}
[(1-\rho_i \phi)^2-\mathcal{R} (1+\rho_i^2 \mathcal{A})]^{\pm\frac{1}{2}}= v^{\pm 1}(1\pm\frac{1}{2 v^2}\sum_n d_n \mathcal{A}^n).\nonumber
\end{equation}

By inserting Eq.(\ref{eqa4}) and Eq.(\ref{eqa6}) into Eq.(\ref{eqa3}), we have
\begin{eqnarray}
&&-4\Omega\sum_{n=1}n(n-1)c_n\mathcal{A}^n-\frac{4}{\mathcal{R}^2}\sum_{n=1}\bigg(b_n+\chi a_n+\frac{1-\chi}{\rho_i}d_n\bigg)\mathcal{A}^{n+1}\sum_{n=1}n(n-1)c_n\mathcal{A}^{n-2}\nonumber\\
&&-\frac{8}{\mathcal{R}}\bigg[\frac{1+\chi+\rho_i(1-\chi)}{\mathcal{R}}-\Omega\bigg]\mathcal{A}^2
\bigg(\sum_{n=1}nc_n\mathcal{A}^{n-1}\bigg)^3\nonumber\\
&&+\frac{4}{v^2\mathcal{R}^2}\sum_{n=1}[b_n+\chi a_n+\rho_i(1-\chi)d_n]\mathcal{A}^{n+2}\bigg(\sum_{n=1}nc_n\mathcal{A}^{n-1}\bigg)^3\nonumber\\
&&+\frac{4}{\mathcal{R}^2}\bigg\{-\frac{1}{2v^2}\sum_{n=1}[b_n-\chi a_n-(1-\chi)d_n]\mathcal{A}^{n+1}
+[1+\chi+\rho_i(1-\chi)]\sum_{n=1}c_n\mathcal{A}^{n+1}\nonumber\\
&&-\frac{1}{2v^2}\sum_{n=1}c_n\mathcal{A}^{n+1}\sum_{n=1}[b_n+\chi a_n+\rho_i(1-\chi)]\mathcal{A}^n\bigg\}\bigg(\sum_{n=1}nc_n\mathcal{A}^{n-1}\bigg)^2\nonumber\\
&&+2\bigg\{\bigg[\frac{1+\chi+\rho_i(1-\chi)}{\mathcal{R}}-2\Omega\bigg]\mathcal{A}-\frac{1}{\mathcal{R}^2}
\sum_{n=1}\bigg(b_n+\chi a_n+\frac{1-\chi}{\rho_i}d_n\bigg)\mathcal{A}^n\nonumber\\
&&-\frac{1}{2v^2\mathcal{R}}\sum_{n=1}[b_n+\chi a_n+\rho_i(1-\chi)d_n]\mathcal{A}^{n+1}\bigg\}\sum_{n=1}nc_n\mathcal{A}^{n-1}\nonumber\\
&&-\frac{1}{\mathcal{R}}\bigg\{[1+\chi+\rho_i(1-\chi)]\sum_{n=1}c_n\mathcal{A}^n-\frac{1}{2v^2}
\bigg[\sum_{n=1}(b_n-\chi a_n-(1-\chi)d_n)\mathcal{A}^n\nonumber\\&&
+\sum_{n=1}c_n\mathcal{A}^n\sum_{n=1}(b_n+\chi a_n+\rho_i(1-\chi))\mathcal{A}^n\bigg]\bigg\}=0, \label{eqa7}
\end{eqnarray}
and by using Eq.(\ref{eqa5}) and Eq.(\ref{eqa7}), we can get the  first two expansion coefficients of $\phi$, i.e., $c_1$ and $c_2$:
\begin{eqnarray}
&&c_1=\frac{\mathcal{R}(1-\chi)(1-\rho_i^2)}{2(4-3\mathcal{R})[1+\chi+\rho_i(1-\chi)]
-8\mathcal{R}(1-\mathcal{R})\Omega},\label{eqa8}\\
&&c_2=c_1\kappa_1/\kappa_2, \label{eqa9}
\end{eqnarray}
where
\begin{eqnarray}
\kappa_1=&&\frac{[(1-3\rho_i^2)c_1-2\mathcal{R}\rho_i^3-\rho_i](1-\chi)-\mathcal{R}(1+\chi)}
{2\mathcal{R}(1-\mathcal{R})}\nonumber\\
&&+\frac{2c_1\{2(1+\chi)c_1+(1-\chi)[2\rho_ic_1-\mathcal{R}(1-\rho_i^2)]\}}{\mathcal{R}^2(1-\mathcal{R})}\nonumber\\
&&+\frac{6[1+\chi+(1-\chi)\rho_i]-8\mathcal{R}\Omega}{\mathcal{R}^2}c_1^2,\label{eqa10}\\
\kappa_2=&&\frac{[1+\chi+(1-\chi)\rho_i](16-15\mathcal{R})}{\mathcal{R}(1-\mathcal{R})}-16\Omega. \label{eqa11}
\end{eqnarray}

As $g(\phi(\mathcal{A}),\mathcal{A})\equiv g(\mathcal{A})$, Eq.(\ref{eq5}) can be written as
\begin{equation}
2\mathcal{A}\left(\frac{d^2\mathcal{A}}{d\xi^2}\right)-\left(\frac{d\mathcal{A}}{d\xi}\right)^2-4\mathcal{A} g(\mathcal{A})=0,\label{eqa12}
\end{equation}
and its first integral is
\begin{equation}
\frac{1}{\mathcal{A}}\left(\frac{d\mathcal{A}}{d\xi}\right)^2-4\int\frac{g(\mathcal{A})}{\mathcal{A}}d\mathcal{A}
=\textup{const}=0.\label{eqa13}
\end{equation}

Expanding $g(\mathcal{A})$ with Eq.(\ref{eqa4}) and Eq.(\ref{eqa6}), then Eq.(\ref{eqa13}) can be reduced to
\begin{equation}
\left(v\frac{d\mathcal{A}}{d\xi}\right)^2=4v^2\left[\frac{1+\chi+\rho_i(1-\chi)}{\mathcal{R}}-\Omega\right]\mathcal{A}^2
-\frac{2}{\mathcal{R}}\sum_{n=1}\frac{b_n+\chi a_n+\rho_i(1-\chi)d_n}{n+1}\mathcal{A}^{n+2}.\label{eqa14}
\end{equation}

With a limit of $n\le2$, Eq.(\ref{eqa14}) becomes
\begin{equation}
\left(v\frac{d\mathcal{A}}{d\xi}\right)^2=\alpha_1\mathcal{A}^2+\alpha_2\mathcal{A}^3+\alpha_3\mathcal{A}^4,\label{eqa15}
\end{equation}
where
\begin{eqnarray}
&&\alpha_1=\frac{4(1-\mathcal{R})}{\mathcal{R}}[1+\chi+\rho_i(1-\chi)-\mathcal{R}\Omega],\nonumber\\
&&\alpha_2=1+\chi+\rho_i^3(1-\chi)-\frac{2(1-\rho_i^2)(1-\chi)}{\mathcal{R}}c_1,\label{eqa16}\\
&&\alpha_3=-\frac{2}{3\mathcal{R}}\{[1+\chi+\rho_i^3(1-\chi)]c_1^2+2(1-\rho_i^2)(1-\chi)c_2\}.\nonumber
\end{eqnarray}
For $\alpha_1>0$, the solution of Eq.(\ref{eqa15}) is got
\begin{equation}
\mathcal{A}=\frac{\beta_+\beta_-\textup{sech}^2(\sigma\xi)}{\beta_+-\beta_-\tanh^2(\sigma\xi)},\label{eqa17}
\end{equation}
where
\begin{eqnarray}
&&\sigma=\frac{\sqrt{\alpha_1}}{2v},\nonumber\\
&&\beta_\pm=\frac{-\alpha_2\pm\sqrt{\alpha_2^2-4\alpha_1\alpha_3}}{2\alpha_3}.\label{eqa18}
\end{eqnarray}
As the argument in Ref.\cite{Bere}, for $R$ satisfying
\begin{equation}
|c_4\mathcal{A}_{max}^4|\le|c_3\mathcal{A}_{max}^3|\ll|c_1\mathcal{A}_{max}+c_2\mathcal{A}_{max}^2|<1,\label{eqa19}
\end{equation}
the series Eq.(\ref{eqa4}) converged so rapidly that the solution Eq.(\ref{eqa17}) is almost the same to the exact one, see Fig.\ref{Fig4}.

Furthermore, from $\alpha_1>0$, we can get
\begin{equation}
0< \mathcal{R}<\frac{1+\chi+\rho_i(1-\chi)}{\Omega},\label{eqa20}
\end{equation}
which is the same region as we have got with nonlinear dynamics method.
So to keep the series expansion method is valid, the range of values for $\mathcal{R}$ must satisfy Eq.(\ref{eqa19}) and Eq.(\ref{eqa20}). Another thing should be noted is that when $(1-\chi)\ll1$ this analytical method is invalid.

Finally it is noted that when $\rho_i=0$, the above solution can be recovered to the case of immobile ion which has been studied in Ref.\cite{Bere}.

\end{appendix}


\begin{thebibliography}{99}\suppressfloats

\bibitem{Pol}
A. I. Akhiezer and R. V. Polovin, Sov. Phys. JETP \textbf{3}, 696 (1956).

\bibitem{Whit}
G. B. Whitham, Linear and Nonlinear Waves, (Wiley, New York, 1974).

\bibitem{Bulanov}
S. V. Bulanov, T. Z. Esirkepov, N. M. Naumova, F. Pegoraro, and V. A. Vshivkov, Phys. Rev. Lett. \textbf{82}, 3440 (1999).

\bibitem{Berezhiani1}
V. I. Berezhiani, V. Skarka, and S. Mahajan, Phys. Rev. E \textbf{48}, 3252 (1993).

\bibitem{D.Farina}
D. Farina and S. V. Bulanov,  Phys. Rev. Lett \textbf{86}, 5289 (2001).

\bibitem{Xie}
B. S. Xie and S. C. Du, Phys. Plasma \textbf{13}, 074504 (2006).

\bibitem{Xie1}
B. S. Xie and S. C. Du, Phys. Scr.\textbf{74}, 638 (2006).

\bibitem{Farina}
D. Farina, S. V. Bulanov,  Phys. Rev. E \textbf{64}, 066401 (2001).

\bibitem{Chen}
H. Chen et al., Phys. Rev. Lett. \textbf{102}, 105001 (2009).

\bibitem{Michel}
F. C. Michel, Rev. Mod. Phys. \textbf{54}, 1 (1982).

\bibitem{Begelman}
M. C. Begelman, R. D. Blandford, and M. D. Rees, Rev. Mod. Phys. \textbf{56}, 255 (1984).

\bibitem{Sturrock}
P. A. Sturrock, Astrophys. J. \textbf{164}, 529 (1971).

\bibitem{Berezhiani}
V. I. Berezhiani, D. D. Tskhakaya, and P. K. Shukla, Phys. Rev. A \textbf{46}, 6608 (1992).

\bibitem{Bere}
V. I. Berezhiani, M. Y. El-Ashry, and U. A. Mofiz, Phys. Rev. E, \textbf{50}, 448(1994).

\bibitem{Berezhiani2}
V. I. Berezhiani and S.M.Mahajan, Phys. Rev. Lett. \textbf{73}, 1110 (1994);
V. I. Berezhiani and S. M. Mahajan, Phys. Rev. E \textbf{52}, 1968 (1995).

\bibitem{Popel}
S. I. Popel, S. V. Vladimirov, and P. K. Shukla, Phys. Plasma \textbf{2}, 716 (1995).

\bibitem{Shukla}
P. K. Shukla, A. A. Mamun, and L. Stenflo, Phys. Scr. \textbf{68}, 295 (2003).

\bibitem{Haque}
Q. Haque and H. Saleem, Phys. Scr. \textbf{69}, 406 (2004).

\bibitem{Kakati}
H. Kakati and K. S. Goswami, Phys. Plasma \textbf{7}, 808 (2000).

\bibitem{Mishra}
M. K. Mishra, R. S. Tiwari, and S. K. Jain, Phys. Rev. E \textbf{76}, 036401 (2007).

\bibitem{Mahmood}
S. Mahmood, H. Ur-Rehman,  Phys. Lett. A \textbf{373}, 2255-2259 (2009).

\bibitem{Lehmann}
G. Lehmann, K. H. Spatschek,  Phys. Rev. E \textbf{83}, 036401 (2011).

\bibitem{Hua}
C. C. Hua, B. S. Xie and K. F. He, Chaos, Solitons and Fractals, \textbf{25}, 1161, (2005).

\bibitem{varma}
R. K. Varma and N. N. Rao, Phys. Lett. A \textbf{79}, 311 (1980).

\bibitem{Rao}
N. N. Rao and R. K. Varma, J. Plasma Phys. \textbf{27}, 95 (1982).

\bibitem{N.N}
N. N. Rao, R. K. Varma, P. K. Shukla, and M. Y. Yu, Phys. Fluids, \textbf{26}, 2488 (1983).

\bibitem{Mofiz}
A. U. Mofiz, U. De Angelis, J. Plasma Phys. \textbf{33}, 107 (1985).

\end{thebibliography}
\end{document}